\documentclass[aps,pre,twocolumn,showpacs,floatfix,superscriptaddress, graphics]{revtex4-1}
\usepackage{amsmath,amssymb,graphicx,color,braket,hyphenat,makeidx,xcolor}
\usepackage[ocgcolorlinks,colorlinks=true,linkcolor=blue,citecolor=red,linktocpage=true]{hyperref}

\usepackage{bm}% bold math
\usepackage{subfigure}
\usepackage{epstopdf}
\usepackage{braket}
\usepackage{enumerate}

\newcommand{\tr}{\mathrm{Tr}}
\newcommand{\E}{{\mathcal E}}
\newcommand{\D}{{\mathcal D}}
\newcommand{\HAT}{}

\begin{document}
\title{Entropy production and fluctuations in a Maxwell's refrigerator with squeezing}
%\subtitle{Do you have a subtitle?\\ If so, write it here}
\author{Gonzalo Manzano}
\email{gmanzano@ucm.es}

\affiliation{Departamento de F\'isica At\'omica, Molecular y Nuclear and GISC, Universidad Complutense de Madrid, 28040 Madrid, Spain} 
\affiliation{Institute for Cross-disciplinary Physics and Complex Systems (UIB-CSIC), Campus Universitat de les Illes Balears, E-07122 Palma de Mallorca, Spain}
\begin{abstract}
 Clarifying the impact of quantumness in the operation and properties of thermal machines represents a major challenge. Here we envisage a toy model acting either as an information-driven 
fridge or as heat-powered information eraser in which coherences can be naturally introduced in by means of squeezed thermal reservoirs. We study the validity of the transient entropy production fluctuation 
theorem in the model with and without squeezing as well as its decomposition into adiabatic and non-adiabatic contributions. Squeezing modifies fluctuations and introduces extra mechanisms 
of entropy exchange. This leads to enhancements in the cooling performance of the refrigerator, and challenging Landauer's bound in the setup.
\end{abstract}

\maketitle
\section{Introduction}
\label{intro}
In 1871 Maxwell proposed his famous thought experiment illustrating the deep link between information and thermodynamics, today known as \textit{Maxwell's demon} \cite{Plenio, Mayurama, ParrondoInfo}.
In the though experiment, a little intelligent being acquires information about the positions and velocities of the molecules of two gases at different temperatures. 
The gases are separated by a rigid wall in containers of equal volume, and the demon can control a tiny door in the wall which can be open or closed, letting fast (hot) particles of the 
cold temperature container pass to the hot temperature container when they approach the wall. This may result in a paradoxical heat current against a temperature gradient without any 
invested work, a situation which challenges the second law of thermodynamics.

Nowadays it is well understood that information modifies the energetic restrictions imposed by the second law \cite{Plenio, Mayurama, ParrondoInfo}, as anticipated by Leo Szilard in 1929 \cite{Szilard}. 
One of the most popular solutions to Maxwell's demon paradox is due to Charles Bennett \cite{Bennett}, who invoked Landauer's erasure principle \cite{Landauer}. Following Landauer, 
logical erasure of a bit of information in a system in contact with a thermal reservoir at temperature $T$ requires a minimum dissipation of heat, $Q_\mathrm{eras} \geq k_B T \ln 2$, 
which is called Landauer's bound. Landauer's principle has been experimentally verified at the single particle level both in classical \cite{Berut} and quantum systems \cite{Peterson}. 
Moreover, both classical and quantum devices acting as Maxwell's demon have been recently proposed \cite{Mandal, MandalII, Strasberg, Chapman, Elouard} and 
implemented in the laboratory in a variety of physical systems \cite{Toyabe, Koski, Roldan, KoskiII, Photonic, Camati, Cottet}.

A major challenge in quantum thermodynamics \cite{PaulReview, AndersReview} is to determine the impact of genuinely quantum effects in the thermodynamic behavior and 
properties of small systems subjected to fluctuations. This is specially interesting in the case of non-ideal environments, such as finite \cite{EspositoCorrelations, Averin, Reeb, Suomela} 
or non-thermal reservoirs \cite{Vaccaro, LutzNeqR, Alicki, EspositoPRX}. In particular, quantum coherence, squeezing and quantum correlations have been shown to improve the ability to extract work 
and the performance of quantum thermal machines \cite{Scully, LutzCorrelations, Li, LutzSqueez, Correa, Hardal, SqzRes, Niedenzu, Segal}. 
However, unveiling the mechanisms responsible of such improvements and bounding their effects represents an open problem, which requires a precise formulation of the second law of 
thermodynamics in such non-equilibrium situations \cite{SqzRes, Segal}. In this context, the development of fluctuation theorems in the quantum regime 
\cite{CampisiFTReview, EspositoFTReview} has provided useful tools to understand the deep relation between dissipation and irreversibility. Indeed fluctuation theorems 
have been very recently extended to the case of a quantum system coupled to non-ideal reservoirs and following arbitrary quantum evolutions \cite{New} (see also Ref. \cite{MHP}).
These new extensions can help to unveil the thermodynamic properties of small thermal machines operating with non-thermal quantum reservoirs, being the squeezed thermal reservoir 
one of the most paradigmatic examples.

Squeezing constitutes a useful resource not only from the perspective of quantum information, with applications in quantum metrology, imaging, computation, and cryptography \cite{Polzik:2008}, 
but also from the perspective of quantum thermodynamics, as its presence in an otherwise thermal reservoir introduces modifications in the second law \cite{SqzRes}. 
The defining feature of squeezing is the introduction of asymmetry between the position and momentum uncertainties \cite{Ficek}. This allows squeezed thermal states to support higher 
energies for the same value of the entropy than Gibbs thermal states, and induces important modifications in the energy fluctuations \cite{Fearn, Kim}. Squeezed thermal states were first 
implemented in the lab for light at microwave frequencies \cite{Yurke}, while nowadays they can be implemented even in massive high-frequency mechanical oscillators 
\cite{Wollman, Pirkkalainen, Rashid}. A proposal for mimicking a squeezed thermal reservoir in a quantum heat engine configuration has been presented for an ion trap configuration 
\cite{LutzSqueez, LutzAtom}. Furthermore, a recent experiment has shown the operation of a nanobeam heat engine coupled to a squeezed thermal reservoir, where enhancements over 
Carnot efficiency has been experimentally checked \cite{Klaers}.

In this article, we analyze the impact of reservoir squeezing in the fluctuations and thermodynamics of a quantum device acting either as an information power fridge (Maxwell's refrigerator) 
or as a heat driven information eraser (Landauer's eraser). A classical model of an autonomous device showing these two regimes has been proposed in Ref.~\cite{Mandal}, and extended to the 
quantum regime in Ref.~\cite{Chapman}. We propose a different model which is reminiscent of the B\"uttiker-Landauer ratchet \cite{Buttiker, Landauer-But}. 
Here the exchange of heat between reservoirs at different temperatures are controlled by a semi-infinite energyless memory system. Our approach consist in analyzing the dynamical evolution 
of this system within the framework of quantum trajectories. This allows us to apply recent results in quantum fluctuation theorems \cite{MHP, New} to the present configuration, which in 
turn provides a precise statement of the second law of thermodynamics in the setup. Comparing the cases in which reservoirs are purely thermal and thermal squeezed, we then obtain explicit 
expressions for the enhancements in the device performance, illustrating how Landauer's principle is modified by means of environmental squeezing.

\section{Setup}
\label{sec:setup}

\begin{figure}[t]
\begin{center}
\includegraphics[width=0.9 \linewidth]{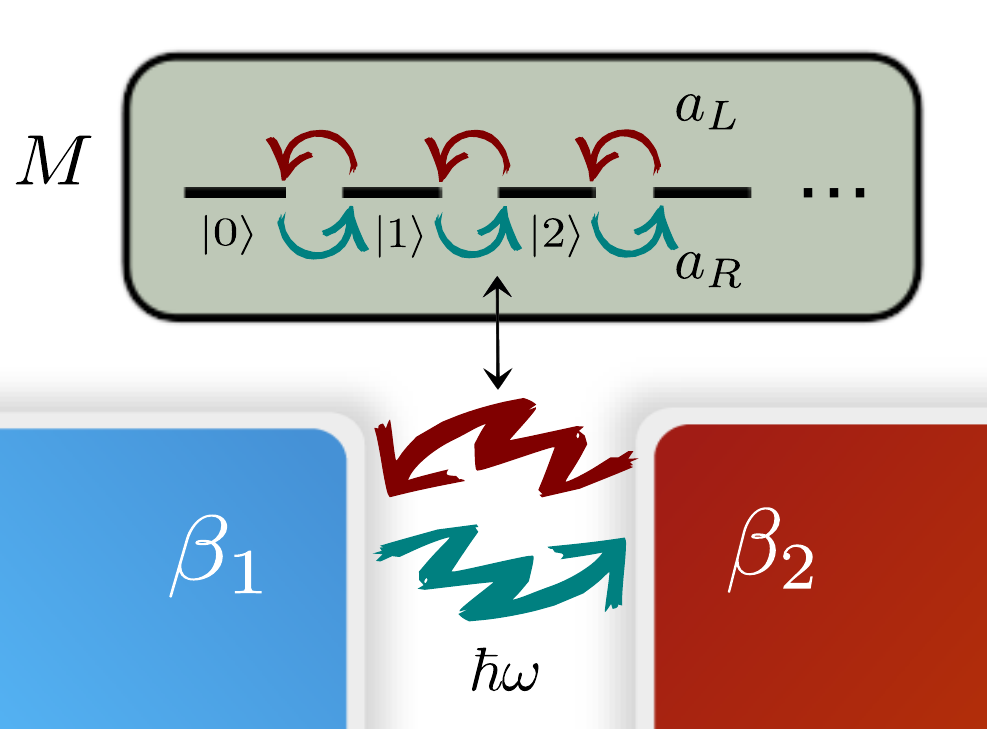}
\caption{Schematic diagram of the setup. Two reservoirs of resonant bosonic modes at different (inverse) temperatures $\beta_1 \geq \beta_2$ are able to exchange 
energy by inducing jumps between the degenerate energy levels of the external memory (M). Each time a quantum $\hbar \omega$ of heat is transferred from the hot (cold) 
to the cold (hot) reservoirs, the memory performs a collective jump to the left (right) as given by the operator $a_L$ ($a_R$).} 
\label{F-Maxwell}
\end{center}
\end{figure}

Consider a small thermal device acting between two resonant bosonic reservoirs at different (inverse) temperatures $\beta_r = 1/k_B T_r$, and an external memory system, $M$, 
in which information can be erased or stored when it is put in contact with the device (see Fig. \ref{F-Maxwell}). The memory is considered here to be a semi-infinite set of 
quantum levels $\{\ket{0}, \ket{1}, ... , \ket{n}, ... \}$ with degenerate energies (conveniently set to zero), and ladder operators $[a_L^{~}, a_L^\dagger] = \mathbb{I}$, producing jumps 
between the degenerate levels to the left ($a_L$) or to the right ($a_R \equiv a_L^\dagger$). The memory $M$ and the reservoir modes weakly couple throughout a three-body interaction term:
\begin{equation}\label{interaction}
 H_I = \hbar g \left( a_L b^{\dagger} c + a_R b c^{\dagger} \right),
\end{equation}
where $g  \ll \omega$, being $\omega$ the natural frequency of the reservoir modes, $H_1 = \hbar \omega b^\dagger b$ and $H_2 = \hbar \omega c^\dagger c$, and where 
$[b,b^\dagger] = \mathbb{I}$ $([c,c^\dagger] = \mathbb{I})$ are the ladder operators of the two reservoir bosonic modes. The above interaction Hamiltonian preserves energy 
and induces jumps on the memory to the left (right) when an energy quantum is transferred from the reservoir $2$ ($1$) to the reservoir $1$ ($2$). The underlying idea of the model is 
to profit from environment induced heat flows throughout the device modes in order to push the state of the memory as much as possible to its leftmost level $\ket{0}$ (Landauer's erasure), 
or alternatively, use the memory in order to induce the desired heat flows between the reservoir against constrains imposed by the environment (Maxwell's refrigerator).

We first consider the case in which both reservoirs are ideal thermal reservoirs in equilibrium at temperatures $\beta_1 \geq \beta_2$. This means that the environmental bosonic modes are assumed 
to be always in a Gibbs state. We are interested in the relaxation dynamics of this model when starting from an arbitrary initial state in the memory. Using standard techniques form open quantum 
system theory \cite{BreuerBook}, one arrives to the following master equation for the dissipative dynamics of the memory density operator, $\rho_t$:
\begin{align}\label{MEq}
\dot{\rho}_{t} = \mathcal{L}(\rho_t) =& \Gamma_\leftarrow \left( a_L \rho_t a_R - \frac{1}{2}\{ a_R a_L , \rho_t \} \right) \nonumber \\
&+ \Gamma_\rightarrow \left( a_R \rho_t a_L - \frac{1}{2}\{ a_L a_R , \rho_t \} \right),
\end{align}
where we neglected Lamb-Stark frequency shifts. The above equation can be straightforwardly checked to be in Lindblad form \cite{Lindblad} by identifying the (two) Lindblad operators 
$L_{\leftarrow} = \sqrt{\Gamma_\leftarrow} ~a_L$ and  $L_{\rightarrow}=\sqrt{\Gamma_\rightarrow} ~a_R$. 
Equation \eqref{MEq} describes incoherent jump processes in the memory to the left at rate $\Gamma_\leftarrow = \gamma (n_{1}^{\rm th} + 1) n_{2}^{\rm th}$, and to the 
right at rate $\Gamma_\rightarrow = \gamma (n_{2}^{\rm th} + 1) n_{1}^{\rm th}$, $\gamma$ being a constant depending on the interaction strength and 
$n_{r}^{\rm th} \equiv (e^{\beta_r \hbar \omega} - 1)^{-1}$. Contact with the thermal reservoirs implies a detailed balance relation between jumps   
\begin{equation}\label{chem}
\Gamma_\leftarrow = e^{\mu} \Gamma_\rightarrow, ~~~~ \mathrm{where}~~~ \mu \equiv (\beta_1 - \beta_2) \hbar \omega \geq 0.
\end{equation}

In the long-time run, the memory system reaches a steady state where jumps to the left and to the right becomes equally probable. This steady state can be obtained from the 
master equation \eqref{MEq} by setting $\mathcal{L}(\pi) = 0$. We obtain
\begin{equation}\label{pi}
\pi = \frac{e^{- \mu N_M}}{Z}, 
\end{equation}
where $N_M = a_L^\dagger a_L^{~}$ is the number operator in the memory, and $Z= \tr[\exp(- \mu N_M)] = (1 - e^{-\mu})^{-1}$. 
The quantity $\mu$ fully determines the steady-state occupation in the degenerate energy levels of the memory, together with its entropy $S(\pi) = \mu \braket{N_M}_{\pi} + \ln Z$, 
being $\braket{N_M}_{\pi} = (e^\mu - 1)^{-1}$. Consequently, the greater the temperature gradient between the reservoirs, the greater $\mu$, the more peaked the distribution in the level leftmost, 
and the lower the entropy of the steady state. On the contrary, if the temperatures of the reservoirs are very similar $\beta_1 \rightarrow \beta_2$, we have $\mu \rightarrow 0$, and the steady 
state of the external system approaches the fully mixed state.

Notice that this simple toy model has all the necessary elements to act as Maxwell's demon, where the heat flows are connected to the informational state of the memory system (regardless of its energy). 
On the one hand, if the initial state of the memory is a low entropic state (in particular if it has lower entropy than $\pi$), we can think about it as an information battery powering a flux of heat against 
the temperature gradient introduced by the reservoirs. This flow is maintained until the memory reaches the steady state $\pi$, and has hence to be replaced. On the other hand, if the initial state is very mixed 
(if it has greater entropy than $\pi$), we can think in our device as a Landauer's eraser, which 
profits from the spontaneous heat flow from the hot to the cold reservoirs to purify the memory state \cite{Mandal}. This is an eraser in the sense that whatever the initial state of the memory, the device 
transforms it into a known ready-to-operate state $\pi$ (which ideally corresponds to an error-free pure state $\ket{0}$), hence erasing any information previously encoded on it.

\section{Quantum trajectories}
\label{sec:trajectories}

We now focus on a stochastic thermodynamical description of the model by using the quantum jump approach \cite{JordanPRA, JordanParrondo, Pekola, Liu, Liu2, Suomela-ad, CampisiHE, Galperin, MHP}. 
Within this approach, the dynamics described by the master equation \eqref{MEq} can be unraveled to construct quantum trajectories $\gamma$ which are sampled according with 
some probabilities $P(\gamma)$ when the environment is monitored \cite{Wiseman}. We start by expanding the dynamics in an infinite set of infinitesimal time evolution steps:
\begin{equation}\label{inf-time-ev}
 \rho_{t + dt} = \left( \mathbb{I} + dt \mathcal{L} \right) \rho_t \equiv \E (\rho_t),
\end{equation}
where $\mathcal{L}$ is given in Eq. \eqref{MEq}, and $\E(\rho_t) = \sum_{k} \HAT{M}_k \rho_t \HAT{M}_k^\dagger$ is a CPTP map fulfilling $\sum_{k} M_k^\dagger M_k = \mathbb{I}$. 
Here the set $\{M_k\}$ with $k = \{0, \leftarrow, \rightarrow \}$ are the Kraus operators of the map. They introduce an interpretation of the dynamics as different random operations 
$\E_k(\rho_t) \equiv M_k \rho_t M_k^\dagger$ occurring with probability $p_k = \tr[M_k^\dagger M_k \rho_t]$. We have
\begin{align}\label{lop}
 M_0 &\equiv ~\mathbb{I} -  \frac{1}{2} \left( \Gamma_\leftarrow a_L^\dagger a_L + \Gamma_\rightarrow a_R^\dagger a_R \right), \\ 
 M_{\leftarrow} &\equiv \sqrt{dt} L_{\leftarrow} = \sqrt{dt ~\Gamma_\leftarrow} ~a_L, \\ \label{lopf}
 M_{\rightarrow} &\equiv \sqrt{dt} L_{\rightarrow} = \sqrt{dt ~\Gamma_\rightarrow} ~a_R,
\end{align}
where we identified the single pair of Lindblad operators introduced before $\{ L_{\leftarrow}, L_{\rightarrow} \}$, and we notice that $\E(\pi) = \pi$.

From Eqs. \eqref{lop}-\eqref{lopf}, it follows that the dynamical evolution is mostly characterized by the operation corresponding to $M_0$, as it occurs with probability of order 1. In contrast, 
operations associated to $M_\leftarrow$ and $M_\rightarrow$ occur with probability $dt$, leading to abrupt jumps in the state of the system. Therefore, between consecutive jumps (e.g. occurring 
at $t$ and $t^\prime$) the evolution is smooth but nonunitary, as given by the effective evolution operator
\begin{align} \label{ueff}
& {U}_{\rm eff} (t^\prime ,t) = \exp \left[ -\frac{i}{\hbar} H_{\rm eff} (t^\prime - t) \right], \\
& H_{\rm eff} = \mathbb{I} - \frac{i \hbar}{2} \left( \Gamma_\leftarrow a_L^\dagger a_L + \Gamma_\rightarrow a_R^\dagger a_R \right).
\end{align}
This is the basis for an alternative description of the dynamics, the so-called direct jump unraveling of the master equation. Within this approach the memory system state no longer evolves according 
to Eqs. \eqref{inf-time-ev} or \eqref{MEq}, but it is given by a stochastic differential equation (the stochastic Schr\"odinger equation) providing the state of the system conditioned on the observed 
outcomes \cite{Wiseman}. 

A trajectory $\gamma$ can then be specified as the result of the following forward process. We start with the system in some arbitrary initial state $\rho_0$ at $t=0$ and perform a projective 
measurement in the $\rho_0$ eigenbasis, selecting an eigenstate $\ket{\psi_n}$ with probability $p_n$. Then by monitoring of the environment, we record the set of different jumps $k_j$ and the 
times $t_j$ where they occurred. We finally introduce a second projective measurement on the system at time $t=\tau$, performed in the $\rho_\tau$ eigenbasis as given by the solution of the master 
equation \eqref{MEq}. This leaves the system in some final state $\ket{\phi_m}$ with associated eigenvalue $p_m^\ast$. A trajectory with $N$ jumps can then be represented by the set 
$\gamma = \{ n, (k_1,t_1), ..., (k_j,t_j), ..., (k_N, t_N), m\}$, where $n$ and $m$ are the outcomes of the initial and final projective measurements \cite{JordanParrondo, New}. 
The probability of such a trajectory is given by
\begin{align}\label{Pfor}
 P(\gamma) =& \tr[\Pi_m^\ast \mathcal{U}_{\tau, t_N} \E_{k_N} \mathcal{U}_{t_N, t_{N-1}} ... ~\E_{k_l}~ \nonumber \\
            &~~~~~     ... ~\mathcal{U}_{t_2, t_1} \E_{k_1} \mathcal{U}_{t_1, 0} (\Pi_n \rho_0 \Pi_n)], ~~~~
\end{align}
with $ \mathcal{U}_{t_{j+1}, t_{j}}(\rho)={U}_{\rm eff} (t_{j+1} ,t_j)\rho\,{U}^\dagger_{\rm eff} (t_{j+1} ,t_j)$ and where for simplicity we considered initial and final rank-1 projectors 
$\Pi_n \equiv \ket{\psi_n} \bra{\psi_n}$ and $\Pi_m^\ast \equiv \ket{\phi_m} \bra{\phi_m}$ respectively.

\subsection{Backward process and total entropy production}

Trajectories $\gamma$ can be associated to twin trajectories in which the sequence of events have been inverted, $\tilde{\gamma} = \{m, (k_N, t_N), ..., (k_j,t_j), ..., (k_1,t_1), n \}$, 
sampled from the time-reversed process. This includes changing the order of initial and final projective measurements, and considering a backward dynamics corresponding to the evolution 
generated from time-inversion of the global dynamics in system \textit{and} environment \cite{JordanParrondo, New}. This backward dynamics can be characterized by a concatenation of CPTP 
maps $\tilde{\E}$ in analogy to Eq. \eqref{inf-time-ev}. We call $\tilde{\E}$ the backward map, which is equipped with a set of Kraus operators $\{ \tilde{M}_k \}$ fulfilling \cite{New}
\begin{equation} \label{eq:backops}
 \tilde{M}_k = e^{-\sigma_k^E/2} ~\Theta M_k^\dagger \Theta^\dagger ~~~ \forall k,
\end{equation}
where $\Theta$ is the time-reversal operator in quantum mechanics, and the set $\{ \sigma_k^E \}$ are the stochastic entropy changes in the environment associated to any operation $k$.
This guarantees the validity of the detailed and integral fluctuation theorems for the total entropy production 
\begin{equation} \label{eq:FTS}
 \frac{P(\gamma)}{\tilde{P}(\tilde{\gamma})} = e^{\Delta_\mathrm{i} s_\mathrm{tot}(\gamma)}, ~~~~ \langle e^{- \Delta_\mathrm{i} s_\mathrm{tot}} \rangle_\gamma = 1,
\end{equation}
$\langle A \rangle_\gamma$ denoting the average of quantity $A(\gamma)$ over all trajectories $\gamma$. Here the total entropy production at the trajectory level is given by \cite{New}
\begin{equation}
\Delta_\mathrm{i} s_\mathrm{tot}(\gamma) \equiv \Delta s_{m,n} + \sum_{j \in \gamma} \sigma^E_{k_j}, 
\end{equation}
consisting in two terms. The first one, $\Delta s_{m n} \equiv -\ln p_m^\ast + \ln p_n$, is the stochastic entropy change in the memory system \cite{Seifert, Sagawa, JordanPRA}, whose average 
yields the corresponding change in von Neumann entropy $\langle \Delta s_{m n} \rangle_\gamma = S(\rho_\tau) - S(\rho_0)$, with $S(\rho)=-\tr[\rho \ln \rho]$. The second one is the collection 
of environmental entropy changes $\{ \sigma_{k_j}^E \}$ due to the different jumps $k_j$ during the trajectory $\gamma$. 

Furthermore, in Eq. \eqref{eq:FTS} we introduced $\tilde{P}(\tilde{\gamma})$ as the probability of obtaining trajectory $\tilde{\gamma}$ in the backward process. It reads
\begin{align}\label{Pbac}
 \tilde{P}(\tilde{\gamma}) =& \tr[ \Theta \Pi_n \Theta^\dagger ~\tilde{\mathcal{U}}_{t_1, 0} \tilde{\E}_{k_1} \tilde{\mathcal{U}}_{t_2, t_1}~... ~\tilde{\E}_{k_l}~  \nonumber \\
&~~~~~...  ~ \tilde{\mathcal{U}}_{t_N, t_{N-1}} \tilde{\E}_{k_N} \tilde{\mathcal{U}}_{\tau, t_N} (\Theta \Pi_m^\ast \rho_{\tau} \Pi_m^\ast \Theta^\dagger)],
\end{align}
where $\tilde{\E}_{k}(\rho) = \tilde{M}_k \rho \tilde{M}_k^\dagger$ are the backward operations associated to the Kraus operators in Eq. \eqref{eq:backops}, and 
$\tilde{\mathcal{U}}_{t^\prime, t}(\rho) = \tilde{U}_\mathrm{eff}(t^\prime, t) \rho \tilde{U}_\mathrm{eff}^\dagger(t^\prime, t)$ represents again a nonunitary smooth evolution 
\begin{equation}
\tilde{U}_{\rm eff} (t' ,t) = \exp \left[ \frac{i}{\hbar} \Theta H_{\rm eff}^\dagger \Theta^\dagger (t^\prime - t) \right], 
\end{equation}
where $H_\mathrm{eff}$ is the same as in Eq. \eqref{ueff}.

Following Ref. \cite{New}, for systems with paired Lindblad operators $\{ L_k ,L_{k^\prime} \}$ such that $L_k = \sqrt{\Gamma_k} L$, $L_k = \sqrt{\Gamma_{k^\prime}} L^\dagger$ for some  
$L$ (as it is the case here), one necessarily has $\sigma_0^E = 0$ and $\sigma_{k}^E = \ln (\Gamma_k/\Gamma_{k^\prime}) = - \sigma_{k^\prime}^E$.  
Therefore, we obtain the following stochastic entropy changes in the reservoirs associated to each Kraus operator in Eqs. \eqref{lop}-\eqref{lopf}
\begin{equation}\label{er}
 \sigma_0^E = 0, ~~~~~ \sigma_{\leftarrow}^E = \mu, ~~~~~ \sigma_{\rightarrow}^E = - \mu.
\end{equation}
When a jump to the left (right) occurs, the entropy in the environment increases (decreases) by $\mu = (\beta_1 - \beta_2) \hbar \omega ~ (-\mu)$, associated to the exchange of a quantum of 
energy $\hbar \omega$ from the hot (cold) to the cold (hot) reservoir. For the backward evolution we then have the operators [Eq. \eqref{eq:backops}]: 
\begin{align} \label{back-kraus}
\tilde{M}_0 &= \Theta M_0^\dagger \Theta^\dagger = M_0, \\ 
\tilde{M}_{\leftarrow} &= \sqrt{dt} \tilde{L}_{\leftarrow} = \Theta \sqrt{dt} L_{\rightarrow} \Theta^\dagger = M_{\rightarrow}, \\  \label{back-krausf}
\tilde{M}_{\rightarrow} &= \sqrt{dt} \tilde{L}_{\rightarrow} = \Theta \sqrt{dt} L_{\leftarrow} \Theta^\dagger = M_{\leftarrow}.  
\end{align}
We obtain that the forward and the backward maps are essentially equal, and operations corresponding to a jump to the left in the forward process transforms into a jump to the right 
in the backward process, and vice-versa.

\subsection{Adiabatic and non-adiabatic entropy production}

% Different versions of the 
% time-reversed process can be introduced \cite{Aurell}, leading to detailed and intergral fluctuation theorems for different versions of the entropy production \cite{MHP, New}. 
% In particular, we are interested in comparing with two kind of time-reverse dynamics, namely, the backward dynamics and the dual-reverse dynamics as introduced in Ref. \cite{New}. 

The fluctuation theorems for the total entropy production in Eq. \eqref{eq:FTS} can be further complemented with independent fluctuation theorems for the so called adiabatic and 
non-adiabatic entropy productions \cite{EspositoFT, JordanParrondo, New}. These two entropy productions appear when discussing the different sources of irreversibility in standard 
thermodynamic configurations \cite{EspositoFT, EspositoFaces}, leading to the split 
\begin{equation} \label{eq:decomposition}
 \Delta_\mathrm{i} s_\mathrm{tot}(\gamma) = \Delta_\mathrm{i} s_\mathrm{ad}(\gamma) + \Delta_\mathrm{i} s_\mathrm{na}(\gamma).
\end{equation}
On the one hand, $\Delta_\mathrm{i} s_\mathrm{na}(\gamma)$ represents the non-adiabatic entropy production, created whenever the state of the system $\rho_t$ is different from the steady 
state of the dynamics $\pi$. For time-dependent Hamiltonians, the average non-adiabatic entropy production $\langle \Delta_\mathrm{i} s_\mathrm{na}(\gamma) \rangle_\gamma$ typically vanishes 
for very slow (quasi-static) driving, since in this situation the state of the system follows the (time-dependent) steady state \cite{EspositoFT, EspositoFaces, New}. 
On the other hand, $\Delta_\mathrm{i} s_\mathrm{ad}$ is the adiabatic entropy production, which appears when the dynamical evolution is driven by nonequilibrium external constrains. 
Indeed, the adiabatic entropy production becomes the only source of entropy production for the case in which the system reaches a non-equilibrium steady state.

In Ref. \cite{New} fluctuation theorems for the adiabatic and non-adiabatic entropy production, as the one in Eq. \eqref{eq:FTS}, have been established for a broad class of quantum evolutions.
They are based on the comparison of the forward process statistics $P(\gamma)$ with the ones in two accessory twin processes: the dual and dual-reverse processes, previously introduced in the 
context of stochastic thermodynamics \cite{SeifertReview, EspositoFT}. These two processes are related between them by inversion of time. That is, the dual process 
generates forward trajectories like $\gamma$, while in the dual-reverse process the sequence of events is inverted and, as a consequence, trajectories $\tilde{\gamma}$ are generated. 
However, in both cases the infinitesimal dynamics are defined by new maps, namely the dual $\D$ and dual-reverse maps $\tilde{\D}$. Their corresponding Kraus operations must fulfill \cite{New} 
(see also Refs. \cite{Crooks, MHP})
\begin{align} \label{eq:detailed}
D_k =  e^{- (\sigma^E_k + \Delta \phi_k)/2} ~ M_k, ~~~~ \tilde{D}_k = e^{\Delta \phi_k/2} ~ \Theta M_k^\dagger \Theta^\dagger.
\end{align}
Here we introduced the quantities $\Delta \phi_k$, representing the change in a very particular observable of the system associated to the operator $M_k$ of the forward process, 
the so-called nonequilibrium potential \cite{MHP}
\begin{equation}
\Phi \equiv - \ln \pi = \sum_i \phi_i \ket{\pi_i} \bra{\pi_i}, 
\end{equation}
$\ket{\pi_i}$ being the eigenstates of the steady state $\pi$. The nonequilibrium potential has long been used in classical irreversible thermodynamics with non-equilibrium 
steady states \cite{Hanggi, Hatano} and in the context of fluctuation-dissipation theorems \cite{Prost} (see also \cite{Hanggi}).
Sufficient conditions for Eqs. \eqref{eq:detailed} to hold are \cite{New, MHP}
\begin{equation} \label{condition}
\tilde{\E}(\Theta \pi \Theta^\dagger) = \Theta \pi \Theta^\dagger, ~~~~ [\Phi, L_k]= \Delta \phi_k~L_k,
\end{equation}
when $\E(\pi) = \pi$ is a positive-definite state. All these conditions hold in the present case as we will shortly see. 

The adiabatic and non-adiabatic entropy production contributions in Eq. \eqref{eq:decomposition} can be then defined by comparing the trajectory probabilities between the original and 
the dual and dual-reverse processes \cite{EspositoFT, Seifert}. They result \cite{New}
\begin{align}
& \Delta_\mathrm{i} s_\mathrm{na} (\gamma) = \Delta s_{n m} - \sum_{j \in \gamma} \Delta \phi_{k_j}, \\
& \Delta_\mathrm{i} s_\mathrm{ad} (\gamma) = \sum_{j \in \gamma} \sigma_{k_j}^E - \sum_{j \in \gamma} \Delta \phi_{k_j},
 \end{align}
which fulfill independent detailed and integral fluctuation theorems as the one in Eq. \eqref{eq:FTS}, provided the conditions in Eq. \eqref{condition} are obeyed.

From the steady state \eqref{pi}, the nonequilibrium potential reads: 
\begin{equation}
\HAT{\Phi} = - \ln \pi = \mu N + \ln Z.
\end{equation}
It is now easy to see that the Kraus operators in Eqs. \eqref{lop}-\eqref{lopf} are related to a unique change in the nonequilibrium potential, that is, 
$[\HAT{\Phi}, L_{k}] = \Delta \phi_{k} L_{k}$ for $k= \{\leftarrow, \rightarrow \}$, with associated potential changes
\begin{equation} \label{tp}
 \Delta \phi_0 = 0, ~~~~  \Delta \phi_{\leftarrow} = - \mu, ~~~~ \Delta \phi_{\rightarrow} =  \mu.
\end{equation}
Furthermore, Eqs.~\eqref{back-kraus}-\eqref{back-krausf} ensure that the map $\tilde{\E}$ has as invariant state $\tilde{\pi} = \Theta \pi \Theta^\dagger$.

Remarkably, by comparing Eqs. \eqref{er} and \eqref{tp}, we see that the changes in the nonequilibrium potential produced by the jumps exactly coincides with the decrease in stochastic 
entropy in the reservoirs, that is $\Delta \phi_{\leftarrow, \rightarrow} = - \sigma_{\leftarrow, \rightarrow}$. 
Therefore, we can conclude that in this case the dual-reverse and backward processes are exactly equal, and hence the dual process is just the original forward one, which implies zero 
adiabatic entropy production, $\Delta_\mathrm{i} s_\mathrm{ad}(\gamma) = 0$. That is, the only contribution to the entropy production is non-adiabatic:
\begin{align}\label{eq:nonadia}
\Delta_{\rm i} s^{\rm na}_\gamma &= \Delta_{\rm i} s_\gamma = \Delta s_{n m} -\sum_{j \in \gamma} \Delta \phi_{k_j}  \nonumber \\ 
&\equiv \Delta s_{n m} - (\beta_1 - \beta_2) q_\gamma,
\end{align}
and fulfills the detailed and integral fluctuation theorems in Eq. \eqref{eq:FTS}. In the last equality of Eq. \eqref{eq:nonadia} we identified the net heat, $q_\gamma$,
flowing from the cold to the hot reservoir during the trajectory $q_{\gamma} = \hbar \omega (n_\rightarrow - n_\leftarrow)$, where $n_{\rightarrow (\leftarrow)}$ is the total number 
of jumps right (left) detected during the trajectory $\gamma$. The absence of adiabatic entropy production can be understood from the fact that in this model, any transfer of heat 
between reservoirs is achieved by means of jumps in the memory, cf. Eq.~(\ref{interaction}). This implies that no heat can flow without modifying the system density operator, 
and hence no entropy can be produced irrespective of the changes in $\rho$, which is the defining feature of the non-adiabatic entropy production. As a consequence, any flow ceases 
in the long-time run, when the steady state $\pi$ is reached, blocking the heat transfer between the reservoirs.

\subsection{Average entropy production rate}

The integral fluctuation theorem in Eq. \eqref{eq:FTS} has as an immediate corollary the positivity of the average entropy production $\langle \Delta_\mathrm{i} s_\mathrm{tot} \rangle \geq 0$.
Furthermore, since the theorem can be applied to any infinitesimal instant of time, we immediately obtain the positivity of the total entropy production rate \cite{New}
\begin{equation}
 \dot{S}_\mathrm{tot}  = \dot{S} - \dot{\sigma}^E \geq 0, 
\end{equation}
where $\dot{S} = -\tr[\dot{\rho}_t \ln \rho_t]$ is the derivative of the von Neumann entropy of the system, and the entropy changes in the environment are given by 
$\dot{\sigma}^E dt = \sum_k \tr[M_k^\dagger M_k \rho_t] \sigma_k^E$.

In this case we obtain:
\begin{equation} \label{2ndlaw}
\dot{S}_{\mathrm{tot}} = \dot{S} - \mu \braket{\dot{N}_M} = \dot{S} - (\beta_1 - \beta_2) \dot{Q} \geq 0,
\end{equation}
where $\dot{Q} = \tr[\dot{\rho}_t \hbar \omega N_M]$ is the heat flow from the cold to the hot reservoir. 
The second-law-like inequality (\ref{2ndlaw}), can be now used to illustrate the two different regimes of operations of the device. Indeed it put bounds on the performance of the two regimes of 
the device at any time of the dynamics, i.e. Landauer's eraser and Maxwell's refrigerator respectively. When $\dot{Q} < 0$, that is, heat flows spontaneously from the hot to the cold reservoir, 
the entropy in the memory system is allowed to decrease, $\dot{S} < 0$, until it compensates the entropy produced by the spontaneous heat flow. Following Eq. \eqref{2ndlaw} The heat dissipated 
in the erasure process is lower bounded by
\begin{equation}\label{Land}
(\beta_1 - \beta_2)|\dot{Q}| \geq ~ |\dot{S}|,
\end{equation}
which is a manifestation of Landauer's principle in our setting. Here the quantity $\beta_1 - \beta_2$ arises as an effective temperature of the memory system determining its steady state.
On the other hand, if $\dot{S} > 0$, now the flux of heat can be inverted against the thermal gradient, $\dot{Q} > 0$ refrigerating the cold reservoir, at the price of entropy production in 
the memory system. The performance of the resulting Maxwell's refrigerator can be analogously bounded
\begin{equation}\label{Mfridge}
 \dot{Q} ~\leq~ \frac{1}{\beta_1 - \beta_2} \dot{S},
\end{equation}
where $\dot{Q}$ is the relevant cooling rate, i.e. the amount of heat per unit time extracted from the cold reservoir.

\section{Squeezed thermal reservoir enhancements}
\label{sec:squeezing}

Once the thermodynamic behavior of the model has been analyzed for the case of ideal thermal reservoirs, we move to the case of non-canonical reservoirs. 
We replace both thermal reservoirs at inverse temperatures $\beta_1$ and $\beta_2$, by squeezed thermal reservoirs at the same temperatures, 
and additional complex parameters $\{\xi_1, \xi_2\}$ characterizing the squeezing, where $\xi_i = r_i e^{i\theta_i}$ for $i=1,2$ \cite{Ficek, BreuerBook}. 

The details on the dissipative dynamics induced by such a quantum reservoir in a bosonic mode, together with the analysis of its thermodynamic power have been reported 
in Ref. \cite{SqzRes}. In the present situation we concern, the master equation in Eq. \eqref{MEq} changes to:
\begin{align}\label{MEq-sqz}
\dot{\rho}_{t} = \mathcal{L}^{\ast}(\rho_{t}) =& \Gamma_{-} \left( R ~\rho_t  R^\dagger - \frac{1}{2}\{ R^\dagger R , \rho_t \} \right) \nonumber \\
&+ \Gamma_{+} \left( R^\dagger \rho_t R - \frac{1}{2}\{ R R^\dagger , \rho_t \} \right),
\end{align}
where $R$ is the ladder operator of a Bogoliubov mode defined by the canonical transformation
\begin{equation}
 R \equiv a_L \cosh(r)  + a_R \sinh(r) ~e^{i \theta} = \mathcal{S}(\xi) a_L \mathcal{S}^\dagger(\xi), 
\end{equation}
$\mathcal{S}(\xi) = \exp(\frac{r}{2}(a_L^2 e^{- i \theta} - a_R^2 e^{i \theta}))$ being the squeezing operator on the memory system. The complex 
parameter is $\xi = r e^{i \theta}$, with $\theta \equiv \theta_1- \theta_2$, and $r \geq 0$. The later depends on all the reservoir temperatures and squeezing 
parameters through the relation
\begin{equation}
\tanh(2r) ~\equiv \frac{2 M_1 M_2}{(N_1 + 1) N_2 + (N_2 + 1) N_1},
\end{equation}
where $M_i \equiv - \sinh(r_i) \cosh(r_i) (2 n_{i}^{\rm th} + 1)$ and $N_i \equiv n_{i}^{\rm th} \cosh(2 r_i) + \sinh(r_i)$ for $i = 1,2$. Notice 
that the above equation is only well defined for the right hand side taking values in between $0$ and $1$ (we recall that $r \geq 0$), to which we will restrict from now on. 
In addition we notice that $r \rightarrow 0$ when either $r_1 \rightarrow 0$ or $r_2 \rightarrow 0$.

The operators $R$ and $R^\dagger$ in the master equation (\ref{MEq-sqz}), promote jumps to the left and to the right, respectively, now between the states of the \textit{squeezed 
basis} of the memory system: $\{ \mathcal{S} \ket{0}, \mathcal{S} \ket{1}, ..., \mathcal{S} \ket{n}, ... \}$. The rates of the jumps are given by 
\begin{equation}
 \Gamma_{\mp} = \frac{\gamma}{2} \left(\delta N \pm (N_2 - N_1) \right),
\end{equation}
where $\delta N \equiv \sqrt{\left( (N_1 + 1)N_2 + (N_2 + 1)N_1 \right)^2 - 4 |M_1 M_2|^2}$. Notice that the rates $\Gamma_{\mp}$  
fulfill again a detailed balance relation for a new parameter $\Gamma_{-} = \Gamma_{+}~e^{\mu_\ast}$, where
\begin{equation}\label{muprime}
\mu_\ast \equiv \ln \left( \frac{N_1 \sinh^2(r) + N_2 \cosh^2(r) + N_1 N_2}{N_1 \cosh^2(r) + N_2 \sinh^2(r) + N_1 N_2} \right),
\end{equation}
which can be both greater or lower than $\mu$ depending on $\{r_1, r_2 \}$ (inside the allowed range). 
In particular, $\mu_\ast \rightarrow \mu$ only when both $r_1 \rightarrow 0$ {\it and} $r_2 \rightarrow 0$.

Crucially, the master equation (\ref{MEq-sqz}) now induces the following steady state reminiscent of the squeezed thermal state
\begin{equation}\label{pis}
 \pi_\ast = \mathcal{S}(\xi) ~\frac{e^{-\mu_\ast N_M}}{Z_\ast}~ \mathcal{S}^\dagger(\xi), 
\end{equation}
with $\mu_\ast$ defined in Eq.~(\ref{muprime}), and $Z_\ast = \tr[e^{-\mu_\ast N_M}] = (1 - e^{-\mu_\ast})^{-1}$.

Following the trajectory formalism, the Kraus operators for the map $\E$ now reads
\begin{align}\label{lops}
 & M_0 = \mathbb{I} -  \frac{1}{2} \left( \Gamma_{-} R^\dagger R + \Gamma_{+} R R^\dagger \right), \\ 
 & M_{-} = \sqrt{dt} L_{-} = \sqrt{dt ~\Gamma_{-}} ~R, \\ 
 & M_{+} = \sqrt{dt} L_{+} = \sqrt{dt ~\Gamma_{+}} ~R^\dagger. 
\end{align}
with the new Lindblad operators, $\{ L_{-}, L_{+} \}$, again suitable to obtain the stochastic entropy changes in the reservoirs
\begin{equation}\label{ers}
 \sigma_0^E = 0, ~~~~~ \sigma_{-}^E = \mu_\ast, ~~~~~ \sigma_{+}^E = - \mu_\ast.
\end{equation}
Notice that the entropy changes in the environment $\sigma_{\mp}^E$, associated to the jumps have no longer a clear interpretation in terms of the 
exchange of an energy quantum between the reservoirs, as in this case both energy and coherence are exchanged between the reservoirs in each jump, producing 
(or annihilating) an entropy quantum $\pm \mu_\ast$ in the whole environment. Nevertheless the fluctuation theorems in Eq. \eqref{eq:FTS} hold, and the 
Kraus operators for the backward map still fulfill the same structure than in the previous case 
\begin{align} \label{back-kraus-s}
  \tilde{M}_0 &= \Theta M_0^\dagger \Theta^\dagger = M_0,  \\ 
  \tilde{M}_{-} &= \sqrt{dt} \tilde{L}_{-} = \Theta \sqrt{dt} L_{+} \Theta^\dagger = M_{+}, \\ 
  \tilde{M}_{+} &= \sqrt{dt} \tilde{L}_{+} = \Theta \sqrt{dt} L_{-} \Theta^\dagger = M_{-}.  
\end{align}

The conditions \eqref{condition} for the adiabatic and non-adiabatic decomposition of the total entropy production also hold in this case. Indeed, from the steady 
state (\ref{pis}) the nonequilibrium potential now reads 
\begin{equation}
\HAT{\Phi} = - \ln \pi_\ast = \mu_\ast \mathcal{S} N_M \mathcal{S}^\dagger + \ln Z_\ast,
\end{equation}
which fulfills $[\HAT{\Phi}, L_{k}] = \Delta \phi_{k} L_{k}$ for $k= \{-, +\}$, being
\begin{equation} \label{tps}
 \Delta \phi_0 = 0, ~~~~  \Delta \phi_{-} = - \mu_\ast, ~~~~ \Delta \phi_{+} =  \mu_\ast.
\end{equation}
Moreover, the map $\tilde{\E}$ has as invariant state $\tilde{\pi}_\ast = \Theta \pi_\ast \Theta^\dagger$ as is required. 

From Eqs.~(\ref{ers}) and (\ref{tps}) we see that also in this case the changes in the nonequilibrium potential produced by the jumps exactly coincides with the decrease of stochastic entropy 
in the reservoirs, $\Delta \phi_{\mp} = - \sigma_{\mp}^E$. Hence the adiabatic entropy production remains zero in the squeezed thermal reservoirs case. This means that when the steady state 
$\pi_\ast$ is reached, no further entropy production is needed in order to maintain it. The average entropy production rate reads in this case
\begin{align} \label{2ndlawSqz}
\dot{S}_{\rm tot} &=  \dot{S} - \mu_\ast \tr[\mathcal{S} N_M \mathcal{S}^\dagger \dot{\rho}_t] \nonumber \\  
                  & = \dot{S} - \frac{\mu_\ast}{\hbar \omega}[ \cosh(2 r) \dot{E} - \sinh(2r) \braket{\dot{\mathcal{A}}_M} ] \geq 0, 
\end{align}
where $\dot{E} = \tr[\hbar \omega N_M \dot{\rho}_t]$ is again the energy flux from the cold to the hot reservoir. Notice that here we refuse from identifying this exchanged energy as heat, since it is not 
directly proportional to the entropy changes in the reservoir. On the other hand
\begin{align}\label{assymetry}
\braket{\dot{\mathcal{A}}_M} &= -\frac{\hbar \omega}{2} \tr[(a_R^2 e^{i \theta} + a_L^2 e^{-i \theta}) \dot{\rho}_t] \nonumber \\
                             &= \frac{\hbar \omega}{2} \tr[(p_{\theta/2}^2 - x_{\theta/2}^2) \dot{\rho}_t], 
\end{align}
is a flow of second-order coherences from the reservoirs. This negentropy flow is responsible of inducing squeezing in the memory state, and, as becomes apparent in the last 
equality of Eq. \eqref{assymetry}, it is quantified by the asymmetry generated in the memory second-order moments of the bosonic quadratures in which squeezing is generated, 
namely, $x_{\theta/2} \equiv (a_R e^{i \theta/2} + a_L e^{-i \theta/2})/\sqrt{2}$ and the conjugate quadrature $p_{\theta/2}$, with $[x_{\theta/2}, p_{\theta/2}] = i\hbar$ \cite{SqzRes}. 
% This assymetry flow obeys an exponential law, $\braket{\dot{\mathcal{A}}_M} = - \gamma (\braket{\mathcal{A}_M} - \braket{\mathcal{A}_M}_{\pi_\ast})$, where the steady state 
% value $\braket{\mathcal{A}_M}_{\pi_\ast} = (\hbar \omega/2) \sinh(2r) \coth(\mu_\ast/2)\geq 0$ is reached in the long time run.

\begin{figure*}[t]
\begin{center}
\includegraphics[width = 1.0 \linewidth]{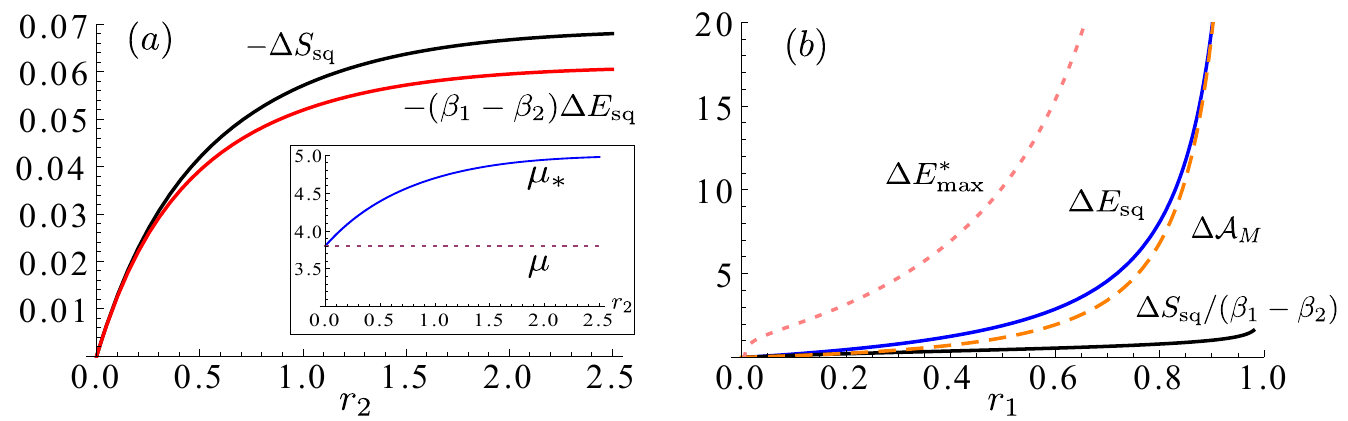}
\caption{(a) Enhancements in the entropy erased $- \Delta S_{\rm sq}$ in the memory, together with the (scaled) energy flow from the hot to the cold 
reservoirs, $-{\Delta E}_{\rm sq}$, as a function of the squeezing parameter $r_2$, in the case of no squeezing in the cold reservoir, $r_1 = 0$. 
In the inset figure we show the change in the parameter $\mu_\ast$ characterizing the steady state of the memory. (b) Enhancements in the energy extracted from the cold 
reservoir (blue line), together with the (scaled) entropy produced in the memory $\Delta S_{\rm sq}$ (black line), the coherences flow to the memory 
$\Delta \mathcal{A}_M$ (orange-dashed line), and the maximum extractable energy from the second-law-like inequality in Eq.~(\ref{2ndlawSqz}), 
$\Delta E_{\rm max}^\ast$ (pink-dotted line) in Eq. \eqref{eq:qmax} as a function of the squeezing parameter $r_1$, when both reservoirs are squeezed ($r_2 = 0.5$). 
All quantities are plotted in units of $\hbar \omega$. In both plots we used $\beta_1 = 5/ \hbar \omega$, and $\beta_2 = 1.2/ \hbar\omega$.} 
\label{Figsqueez}
\end{center}
\end{figure*}

Comparing Eqs.~(\ref{2ndlaw}) and (\ref{2ndlawSqz}) we see two main effects introduced by reservoir squeezing. The first one is the appearance of the parameter $\mu_\ast$ instead of $\mu$. 
This implies a modification in the interplay between the energy transferred between reservoirs and the entropy change in the memory, without any variation of temperatures. 
The second one is the appearance of the extra entropy flow related to the second-order coherences in the memory system, namely $\braket{\dot{\mathcal{A}}_M}$ in Eq. \eqref{assymetry}. 
The later may induce an extra reduction (or increase) of the memory entropy which is independent of the energy exchanged between the reservoirs. Both effects have a deep impact in the performance 
of our device, as we will shortly see (see Fig. \ref{Figsqueez}). To make the discussion more precise, we will look at the device operation when the memory system starts in the state $\pi$ and is 
then connected to the device with the squeezed thermal reservoirs until it reaches the steady state $\pi_\ast$. The extra increase in entropy and energy pumped from the cold to the hot 
reservoirs due to reservoir squeezing read
\begin{eqnarray}\label{enhace}
&\Delta S_{\rm sq} \equiv S(\pi_\ast) - S(\pi) = \frac{\mu_\ast}{e^{\mu_\ast}- 1} - \frac{\mu}{e^\mu -1} + \ln \frac{1 - e^{- \mu}}{1 - e^{-\mu_\ast}},  \\ \label{enhace2}
&\Delta E_{\rm sq} \equiv \hbar \omega \left( \braket{N_M}_{\pi_\ast} - \braket{N_M}_{\pi} \right) = \frac{\hbar \omega}{e^{\mu_\ast}- 1} - \frac{\hbar \omega}{e^\mu -1}. 
\end{eqnarray}
Furthermore, we obtain that the asymmetry induced in the memory system during the process is just
\begin{equation}
\Delta \mathcal{A}_M =  \langle \mathcal{A} \rangle_{\pi^\ast} = \hbar \omega \sinh(2r) \left[(e^{\mu_\ast} - 1)^{-1} + 1/2 \right].
\end{equation}

In Figure \ref{Figsqueez}(a) we show the ``breaking'' of the Landauer's bound in Eq.~(\ref{Land}) when the device acts as an information eraser, $\Delta S_\mathrm{sq} < 0$. Here we just consider squeezing in the hot reservoir, 
$r_2 > 0$, while the cold one remains in a thermal state, $r_1 = 0$. In this case we have that the global squeezing amplitude is $r= 0$, meaning that the steady state in Eq.~(\ref{pis}) is no longer 
squeezed, but just the Gibbs state $\pi_\ast = \exp(\mu_\ast N_M)/Z_\ast$. Nevertheless, notice that this corresponds to a lower entropic state than $\pi$ in Eq.~(\ref{pi}) 
since $\mu_\ast \geq \mu$ (see inset). As a consequence, we have that introducing squeezing only in the cold thermal reservoir we can erase a greater amount of entropy in the memory. This enhanced erasure indeed overcomes the bound (\ref{Land}), 
at the cost of inducing a greater amount of energy transferred from the hot to the cold reservoir, represented by the quantity $(\beta_1 - \beta_2) \Delta E_\mathrm{sq} < 0$.

On the other hand, in Fig. \ref{Figsqueez}(b), we show the performance of the device when acting as a Maxwell's refrigerator. Here we analyze the regime in which both reservoirs contain squeezing, $r_1, r_2 > 0$. 
In this case the final state of the memory is squeezed, $r>0$, enabling extra energy extraction from the cold reservoir, $\Delta E_\mathrm{sq}$. The asymmetry induced in the memory state, as quantified by $\Delta \mathcal{A}_M$, 
together with the modification of the parameter $\mu_\ast$, are responsible of allowing refrigeration on top of the entropy produced in the memory, that is 
$\Delta S_\mathrm{sq}/(\beta_1 - \beta_2)$, overcoming again the bound in Eq.~(\ref{Mfridge}) for the thermal reservoirs case. Nevertheless, following Eq. \eqref{2ndlawSqz} the 
energy that can be extracted in such process never surpass the bound $\Delta E_\mathrm{sq} \leq \Delta E_\mathrm{max}^\ast$, with
\begin{equation} \label{eq:qmax}
\Delta E_\mathrm{max}^\ast \equiv \frac{\hbar \omega}{\mu_\ast} \mathrm{sech}(2 r) \Delta S_\mathrm{sq} + \tanh(2r) \Delta \mathcal{A}_M.
\end{equation}

\section{Conclusions}  
\label{sec:conclusions}

In this paper, we applied recent results on quantum fluctuation theorems \cite{New,MHP} to analyze the thermodynamic impact of squeezing in a Maxwell's demon setup, where a pure informational 
component exist. We first explored the simpler case in which a memory system controls the heat flow between thermal reservoirs at different temperatures. We studied the total entropy production 
in the model both at the level of fluctuations and averages, using a quantum jump approach. We checked the existence of a general fluctuation theorem for the total entropy production in the setup, 
as well as the conditions for its split into adiabatic and non-adiabatic contributions. This analysis allowed the characterization of the device performance in its principal regimes of operation 
in Eqs. \eqref{Land} and \eqref{Mfridge}. 

As a second step, we replaced the regular thermal reservoirs by squeezed thermal reservoirs. In this case the entropy changes in the environment are produced by the exchange of both energy and 
coherence between the nonequilibrium reservoirs. This exchange mechanism is also responsible for inducing squeezing in the memory system at the steady state. The fluctuation theorem for the 
total entropy production holds as well in this case, but the fundamental processes associated to its increase are no longer simple jumps associated to the exchange of energy quanta between reservoirs. 
The total entropy production rate is therefore also modified, now including a term proportional to the asymmetry induced by the memory quadratures [Eq.~(\ref{2ndlawSqz})]. 

The enhancements in the performance of the device due to the squeezing in the reservoirs can be quantified in this model for situations of specific interest [Eqs. \eqref{enhace}-\eqref{enhace2}]. 
As an example we explored the two different situations represented in Fig. \ref{Figsqueez}. These include Landauer's erasure at a lower energetic cost when just one of the 
two thermal reservoirs are squeezed. The second example shows how improvements in the cooling rate of the Maxwell's refrigerator can be achieved when squeezing is present in both reservoirs.
The results presented here pave the way for improved models of quantum devices acting as Maxwell's demon while profiting from quantum thermodynamical resources like squeezing.

%\begin{acknowledgements}
%The authors acknowledge funding from MINECO (grant FIS2014-52486-R). G. M. acknowledges support from FPI grant no. BES-2012-054025. 
%This work has been partially supported by COST Action MP1209 ``Thermodynamics in the quantum regime''.
%\end{acknowledgements}

% 
% \begin{figure}
% % Use the relevant command for your figure-insertion program
% % to insert the figure file.
% % For example, with the option graphics use
% %\resizebox{0.75\columnwidth}{!}{%
% %  \includegraphics{fig1.eps} }
% \caption{Please write your figure caption here.}
% \label{fig:1}       % Give a unique label
% \end{figure}
% %
% % For tables use
% \begin{table}
% \caption{Please write your table caption here.}
% \label{tab:1}       % Give a unique label
% % For LaTeX tables use
% \begin{tabular}{lll}
% \hline\noalign{\smallskip}
% first & second & third  \\
% \noalign{\smallskip}\hline\noalign{\smallskip}
% number & number & number \\
% number & number & number \\
% \noalign{\smallskip}\hline
% \end{tabular}
% \end{table}
% %

\section*{Acknowledgements}

It is a pleasure to thank Juan M. R. Parrondo for useful comments and discussions. G. M. acknowledge funding from MINECO 
(Grants No. FIS2014-52486-R, No. FIS2014-60343-P and FPI Grant No. BES-2012-054025). This work has been partially 
supported by COST Action MP1209 ``Thermodynamics in the quantum regime''.

\section*{Author contributions}

G. M. planned the research, performed the calculations, analyzed the results and wrote the manuscript.

%\bibliography{qft.bib}

\end{document}